\documentclass[twocolumn,pra,aps,superscriptaddress]{revtex4-2}
\usepackage[english]{babel}
\usepackage{amssymb}
\usepackage{bbold}
\usepackage{amsmath}
\usepackage{graphicx}
\usepackage[colorlinks=true, allcolors=blue]{hyperref}
\newcommand{\bla}{bla\\bla\\bla\\bla\\bla}

\begin{document}
\title{Irreversibility  in an optical parametric driven optomechanical system}
\author{Obinna Abah}
\email{obinna.abah@newcastle.ac.uk}
\affiliation{ School of Mathematics, Statistics, and Physics, Newcastle University, Newcastle upon Tyne, NE1 7RU, United Kingdom}

\author{Collins O. Edet}
\affiliation{Institute of Engineering Mathematics, Universiti Malaysia Perlis, 02600 Arau, Perlis, Malaysia. }
\author{Norshamsuri Ali}
\affiliation{Advanced Communication Engineering (ACE) Centre of Excellence, Universiti Malaysia Perlis, 01000 Kangar, Perlis, Malaysia}
\author{Berihu Teklu}
\affiliation{Department of Mathematics, Khalifa University, Abu Dhabi 127788, United Arab Emirates}
\affiliation{Center for Cyber-Physical Systems (C2PS), Khalifa University, 127788 Abu Dhabi, UAE}
\author{Muhammad Asjad}
\email{asjad\_qau@yahoo.com}
\affiliation{Department of Mathematics, Khalifa University, Abu Dhabi 127788, United Arab Emirates}

\begin{abstract}
We investigate the role of nonlinearity via optical parametric oscillator on the entropy production rate and quantum correlations in a hybrid optomechanical system. Specifically, we derive the modified entropy production rate of  an optical parametric oscillator placed in the optomechanical cavity which is well described by the two-mode Gaussian state. We find a dramatic deviation in the irreversibility and quantum mutual information for small detuning. Our analysis shows that the system irreversibility can be reduced by choosing the appropriate phase of the self-induced nonlinearity. We further demonstrate that the nonlinearity effect persist for a reasonable range of cavity decay rate.
\end{abstract}

\maketitle
\section{Introduction}
The performances of thermal heat machines were successfully analyzed within the established framework of classical thermodynamics \cite{Callen1985} and played prominent role during the industrial revolution. In the last decades, thermodynamics has been extended to classical small devices/systems operating far from equilibrium by considering the fluctuations via stochastic thermodynamics \cite{Jarzynski2011,Seifert2012}. In view of harnessing the promises of quantum technologies, there has been a tremendous interest in the thermodynamical analysis of devices operating in quantum regime \cite{Abah2014EPL,GooldPRL,DelCampo2013,Correa2014}. In addition, with the advancement in fabrication technology, various experiments studying nonequilibrium thermodynamics in the quantum regime have been realized \cite{Rossnagel2016,Klaers2017,Sheng2021}. Recently, the irreversible entropy production dynamics in mesoscopic quantum systems have been experimentally measured in two different driven-dissipative  quantum systems realized by coupling bosonic systems to high-finesse cavities \cite{Brunelli2018}. These two experimental platforms are a cavity-optomechanical device, and a Bose-Einstein condensate (BEC) with cavity-mediated long-range inter-actions \cite{Mottl2012,Aspelmeyer2014}.

Hybrid quantum systems exploit different physical components with complementary functionalities for efficient multi-tasking tasks \cite{Kurizki2015}. They catalyse novel fundamental research in quantum mechanics, condensed matter physics, and mesoscopic physics by providing platforms to investigate various phenomena  at the quantum regimes. These systems are also playing a prominent role in realizing the novel range of applications in quantum technologies, including quantum metrology \cite{Geraci2010}, quantum communication \cite{Xiang2017}, as quantum transducers \cite{Stannigel2011,Bagci2014}, for fundamental tests of quantum mechanics \cite{Marshall2003, Romero-Isart2011,Vinante2017} or realizing quantum thermal machines \cite{Zhang2014PRL,Hardal2015,Myers2022}.
However, hybrid cavity optomechanical systems have attracted a lot of attraction due to their integration versatility, promising reliable quantum controllability,  and long coherent time \cite{Zhang2018}. The systems provide a strong analogy between quantum optomechanics, nonlinear optics, and mapping optical effects. Moreover, the degenerate optical parametric oscillator, due to a second-order nonlinearity of optical crystals, placed inside a dissipative cavity optomechanical has been proposed to considerably enhance the entanglement \cite{Pan2023}, mechanical squeezing \cite{Huang2017,Huang2020}, the cooling of the micromechanical mirror \cite{asjad16, Huang2009, Asjad19}, force sensing of the system \cite{He2022} and improve the precision of position detection \cite{Peano2015}. In addition, the  optical parametric oscillator (phase-sensitive amplifier) is offering a promising applications in quantum communication \cite{Adnane2019, Rosati16}  and quantum sensing \cite{Notarnicola2022}. On the otherhand, non-linearity has been shown to be a useful resource for generating non-classical quantum states \cite{katz2007, teklu15, Albarelli16} and a measure of the non-linearity of a quantum oscillator has been proposed, which can quantify the non-lineaity of the oscillator \cite{Paris14}.

In this paper, we study the irreversibility generated in the stationary state of a  nonlinear crystal optical parametric oscillator placed inside the optomechanical cavity. We have  demonstrate that the entropy production rate and the corresponding quantum correlations of the optomechanical setup are modified by this self-induced nonlinearity. We show that the irreversiblity in the system is enhance via squeezing generated from the nonlinear medium but it can be reduced for some specific choise of nonlinear interaction phase. In addition, we analyze the role of self induced nonlinearity on the quantum correlations.

The rest of the paper is structured as follows. In Section \ref{model} we describe the full theoretical model Hamiltonian of  the hybrid optomechanical setup and then proceed to obtain the equations of motion. Following the usual procedure, we linearize the dynamics, where we focus explictly on Gaussian states. Section \ref{results} present the results and discussions of the entropy production rate and the quantum correlations of  model Hamiltonian. First, in Section \ref{entropy} we present the analysis of the entropy production rate of nonlinear hybrid setup while the Section \ref{correlations} details the behaviour of quantum mutual information. Finally, we conclude in Section \ref{conclusions}.

\begin{figure}[!t]
\includegraphics[width=0.95\columnwidth]{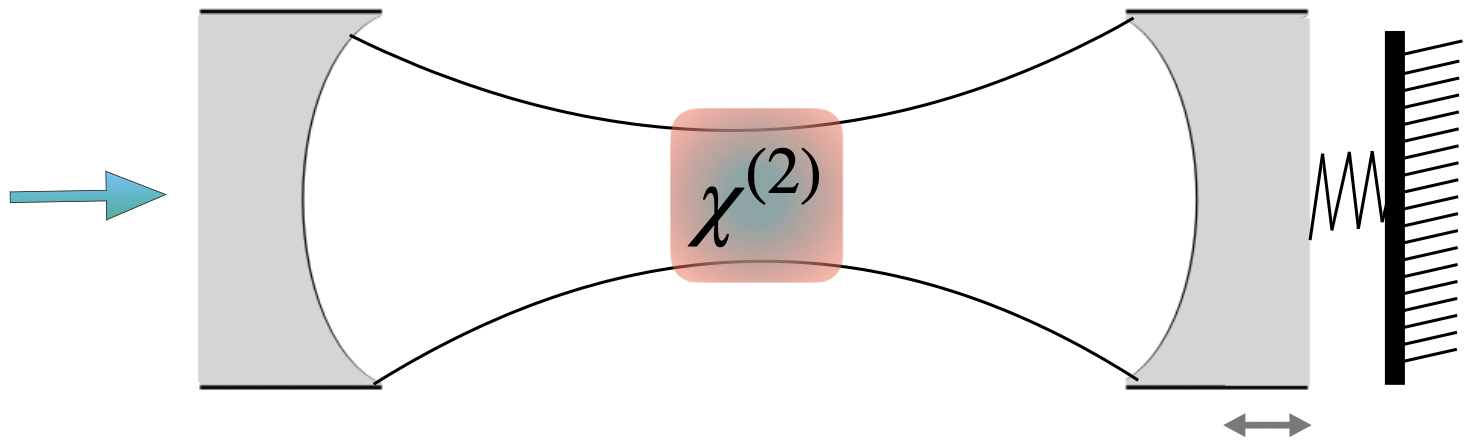} 
\caption{Schematic representation of the optomechanical system  with driven optical parametric oscillator. The optical parametric oscillator (OPO) is sandwiched between the two mirrors in the Fabry-Perot cavity. One of the cavity mirror (right) which is attached to the mechanical oscillator is movable, such that the resonant cavity frequencies is modulated by the mechanical vibrations. The left side mirror is fixed while the optical cavity is pumped/driven by a laser.}          
\label{modelrep}
\end{figure}

\section{Model}
\label{model}
We consider an optical parametric oscillator placed in the optomechanical  cavity, see Fig.~\ref{modelrep}.
The cavity is driven by a laser with frequency $\omega_L$ at rate $\eta$ through one of its end-mirror. The movable cavity mirror is controlled by the mechanical resonator vibrations at frequency $\omega_b$, which modulate the cavity resonance frequencies.  The Hamiltonian of the system in a rotating frame  at the frequency $\omega_L$ of the pump field reads
\begin{equation}
    H=\Delta_a \hat{a}^\dagger \hat{a} + \xi \hat{a}^{\dagger^2} \hat{a}^2 + \omega_b \hat{b}^\dagger \hat{b}+ g \hat{a}^\dagger \hat{a} (\hat{b}+\hat{b}^\dagger) - i\hbar \left(\eta^* \hat{a} - \eta \hat{a}^\dagger\right),
\end{equation}
where $\Delta_a\!=\!\left(\omega_C - \omega_L\right)$ is the cavity detuning with $\omega_C$ is the cavity frequency, $\hat{a}$($\hat{a}^\dagger$) is the cavity field's annihilation (creation) operator and $\xi$ is the strength of nonlinear interaction. The vibrational mode annihilation and creation operators are denoted by $b$ and $b^\dagger$ respectively. The term $g\!=\!\sqrt{\hbar/M\omega_b}\,\omega_C/L$ is the optomechanical coupling between the bare cavity and the mechanics;  $L$ is the cavity length in the absence of the cavity field and $M$ is the mass of the mechanical resonator. The laser rate $\eta=|\eta|e^{i\theta}$ with  $|\eta|=\sqrt{2\kappa \mathcal{R}/\hbar\omega_L}$ ($\mathcal{R}$ is the laser power and $\kappa$ is the cavity decay rate) and $\theta$ is the phase of the deriving lase field. 
Considering the dissipation of mechanical and cavity modes and the corresponding fluctuating noise terms, the quantum Langevin equation for the system described by the Hamiltonian can be written as
\begin{eqnarray}
\Dot{\hat{a}} &=&-\left(\kappa + i\Delta_a \right) \hat{a}-ig\left(\hat{b} + \hat{b}^{\dagger}\right) \hat{a} - 2i \xi \hat{a}^{\dagger} \hat{a}^2+\eta \nonumber \\
&+&\sqrt{2 \kappa} \hat{a}_{in}, \nonumber \\
 \Dot{\hat{b}}&=&-( \gamma + i\omega_{b})\hat{b}-ig \hat{a}^{\dagger} \hat{a} + \sqrt{2 \gamma} \hat{b}_{in}, \label{eq2}
 \end{eqnarray}
where $\gamma$ is the damping rate of mechanical resonator, $\hat{a}_{in}$ is the \textit{zero} mean, \textit{i.e} $\langle \hat{a}_{in}\rangle\!=\!0$, input noise operator for optical mode with only non-zero correlation $\langle \hat{a}_{in}(t) \hat{a}_{in}^{\dagger}(t^{\prime})\rangle\!=\!\delta(t-t^{\prime})$, 
while $\hat{b}_{in}$ is the input noise operator with \textit{zero} mean, $\langle \hat{b}_{in}\rangle\!=\!0$, associated with the mechanical oscillator and described by the correlation function $\langle \hat{b}_{in}(t) \hat{b}_{in}^{\dagger}(t^{\prime}) \rangle=(n_{b} +1)\delta(t-t^{\prime})$ with $n_{b}\!=\!\left(e^{(\hbar \omega_{b}/K_{B}T)}-1\right)^{-1}$ is the mean thermal occupation number of the mechanical mode at the temperature $T$.

To linearize the nonlinear set of equations, Eq. (\ref{eq2}), the quantum operators can be expanded around their respective classical mean values as  =: $a=a_s+\delta a$ and $b=b_s+\delta b$, where $\delta a$ and $\delta b$ are small quantum fluctuations around the mean fields $a_s$ and $b_s$. The steady-state mean-field values are obtained as follows
\begin{eqnarray}
a_{s}  & = & \frac{\eta } {\kappa+ i\Tilde{\Delta}_{a}}, \qquad
b_{s} =  -\frac{g	\left|a_{s}\right|^2}{\gamma+i \omega_{b}}, \\ \nonumber 
\end{eqnarray}
where $\Tilde{\Delta}_{a}=\Delta_{a}+2 g \text{Re} b_{s} + 2\xi \left|a_{s} \right|^{2}$ is the effective detuning by including the self-Kerr induced frequency shifts. Basically, these frequency shifts are minute, i.e., $\left| \Tilde{\Delta}_{a}-\Delta_{a}\right| \ll \Delta_{a} \approx \omega_{b}$. Consequently, from this point forward, we can assume $\Tilde{\Delta}_{a}  \simeq \Delta_{a}$. The related linearized quantum Langevin equations defining the quantum fluctuations dynamics are given by:
\begin{eqnarray}
\delta \Dot{\hat{a}} &=&-\left(\kappa + i\Delta_{a} \right) \delta\hat{a} - ig \left(\delta\hat{b} + \delta \hat{b}^{\dagger}\right) 
- \xi \delta\hat{a}^{\dagger}+\sqrt{2 \kappa} \delta\hat{a}_{in},\nonumber\\
\delta\Dot{\hat{b}} &=&-\left(\gamma+i\omega_{b} \right) \delta\hat{b}-i\left( {G}^{*}\delta\hat{a} + G \delta\hat{a}^{\dagger}\right) + \sqrt{2 \gamma} \delta\hat{b}_{in},
\end{eqnarray}
where ${G}=g \langle a_{s}\rangle$ is the effective optomechanical coupling strength, $\chi \equiv -2i \xi a_{s}^{2}$ represents the effective nonlinear interaction, with the amplitude $\lvert \chi \rvert$ and $\phi = \tan^{-1}[\text{Im}\chi/\text{Re}\chi]$. 

To linearize the equation of motion, which ensures that any initial Gaussian state will remain such at any instant of time \cite{Olivares2012}, we defined the optical quadrature operators $\delta x_a\!=\!(\delta\hat{a}+\delta\hat{a}^\dagger)/\sqrt{2}$ and $\delta p_a\!=\!(\delta\hat{a}-\delta\hat{a}^\dagger)/i\sqrt{2}$. Similarly, the quadratures of the cavity mode are $\delta x_b\!=\!(\delta\hat{b}+\delta \hat{b}^\dagger)/\sqrt{2}$ and $\delta p_b\!=\!(\delta \hat{b}-\delta \hat{b}^\dagger)/i\sqrt{2}$. Likewise, the fluctuation operators $x_{j,\text{in}}$ and $p_{j,\text{in}}$ ($j\!=\!a,b$) are defined in similar fashion. Now, the quantum Langevin equations for the quadratures can be written in the compact matrix form as
\begin{equation}
\dot{\textbf{R}}(t) = A \textbf{R}(t) + \textbf{R}_\text{in}, 
\end{equation}
where the quadratures vector $\textbf{R}(t)\!=\!(\delta x_a, \delta p_a,\delta x_b, \delta p_b)^T$ and the noises vector $\textbf{R}_\text{in}\!=\!(\sqrt{2 \kappa} \, x_{a,\text{in}},\sqrt{2 \kappa}\, p_{a,\text{in}},\sqrt{2 \gamma} \, \delta x_{b,\text{in}}, \sqrt{2 \gamma}\,  \delta p_{b,\text{in}})^T$. The corresponding drift matrix $A$ can be explicitly determined and depends on the set of parameters characterizing the dynamics of the two-mode system; it reads
\begin{equation}
    A= \begin{pmatrix}
    -\kappa+\chi \cos(\phi) & \Delta_a+\chi \sin(\phi) & 0 & 0\\
    -\Delta_a + \chi \sin(\phi) &  -\kappa - \chi \cos(\phi) & G & 0\\
    0 & 0 & -\gamma &  \omega_b\\
    G & 0 & -\omega_b & -\gamma
    \end{pmatrix}.
\end{equation}
The system dynamical equations should be stable in order for a steady state to exist and the stability condition for system can be formalized in terms of the Routh-Hurwitz criterion~\cite{DeJesus1987}, which we employed in our characterization of the dynamics. This is achieved if the real part of the spectrum of the drift matrix $A$ is negative, i.e all the eigenvalues of the drift matrix $A$ have negative real parts. 
\section{Results and discussions}
\label{results}
In this section, we now proceed to analyze the irreversible entropy production rate and the quantum correlation profiles of our setup consisting of an optical parametric oscillator inside the dissipative-driven optomechanical cavity. We will focus on the how these physical quantities are affected by the presence of the OPO when the system reaches its stationary state.
\subsection{Entropy production and correlations matrix}
\label{entropy}
In the quantum domain,  the problem of calculating the entropy production of a quantum system is formulated in terms of the quantum master equations~\cite{Lindblad1975a,Schnakenberg1976,Spohn1978,BreuerBook}, quantum trajectories \cite{Leggio2013} and fluctuation theorems~\cite{Jarzynski1997,Crooks1998} among others. 
In recent, a  formulation for the characterization of irreversible entropy production of  quantum  systems interacting with nonequilibrium reservoirs which combines quantum phase-space methods and the Fokker-Planck equation, has been put forward~\cite{Brunelli2016ArXiv,Santos2017,Santos2018,Zicari2020}. This framework has been employed to experimentally measure and characterize the irreversible entropy production rates of bosonic quantum systems in two platforms- a micromechanical resonator and a Bose-Einstein condensate~ \cite{Brunelli2018}. For the optomechanical system, the cooling of the mechanical resonator is reflected in the entropy production rates. Shahidani and Rafiee have studied the role of self-correlation on irreversible thermodynamics in a parametrically driven-dissipative system~\cite{Shahidani2022}.

Here, we follow the framework that characterizes the entropy production as the correlation between a system and a reservoir~\cite{Brunelli2016ArXiv,Esposito2010b}. Due to the linearized dynamics of the fluctuations and since all the quantum noise terms are Gaussian, the resulting steady state of the system is a continuous-variable Gaussian state which can be fully characterized by the $4 \times 4$ stationary correlation matrix (CM) $\mathcal{V}$, with components $\mathcal{V}_{i j}  =\langle \delta u_i(\infty)\delta u_j(\infty) + \delta u_j(\infty) \delta u_i(\infty)\rangle/2$. The elements of the quantum CM must satisfy the uncertainty relation $\mathcal{V} + i\Omega \ge 0$, where $\omega_{ij}$ are the elements of the symplectic matrix given by the Heisenberg uncertainty principle ($[u_i,u_j]=i\Omega_{ij}$) \cite{Serafini2017}. We assume that the first moments are null, which can be achieved by choosing a suitable displacement in the phase space.
The equation of motion for the covariance matrix as
\begin{equation}
    \dot{\mathcal{V}} =A\mathcal{V} +\mathcal{V} A^T + D,
\end{equation}
where $D\!=\!\text{diag}\{\kappa,\kappa,\gamma (2n_b+1),\gamma (2n_b+1)\}$ is the diffusion matrix. Considering that the two reservoirs are prepared at different temperatures, this leads to the breaking of the detailed balance and takes the system to nonequilibrium state. When the system is assumed to be stable, we obtain the Lyapunov equation for the nonequilibrium steady state covariance matrix $A \mathcal{V}^s + \mathcal{V}^s A^T = -D$.

The open dynamics of the joint system can be described in terms of Fokker-Planck equations based on the Wigner function of the system. Following the approach recently put forward, the steady-state entropy production rate $\Pi_s$ is given by \cite{Brunelli2016ArXiv,Brunelli2018},
\begin{eqnarray}
\Pi_s &=& 2 \mathrm{Tr} \left((A^\text{irr})^T D^{-1} A^\text{irr}\, \mathcal{V}^s\right) + \mathrm{Tr} \left(A^\text{irr} \right) \nonumber\\
&=& 2\kappa \left( \mathcal{V}^s_{11}+\mathcal{V}^s_{22}-1 \right) +  2\gamma \left( \frac{\mathcal{V}^s_{33} +\mathcal{V}^s_{44}}{2n_b +1}-1 \right) , \,\, \nonumber \\
&=& \mu_a + \mu_b,
\end{eqnarray}
where $A^\text{irr}\!=\!\text{diag}\{-\kappa,-\kappa, -\gamma,-\gamma\}$ and $\mu_a$ ($\mu_b$) corresponds to the contributions to $\Pi_s$ from the cavity (mechanical) mode respectively. When the system is in the equilibrium state, we have $\mathcal{V}^s_{11}+\mathcal{V}^s_{22} =  1$, $\mathcal{V}^s_{33}+\mathcal{V}^s_{44} = 2n_b + 1$, and hence, $\Pi_s\!=\! 0$. From the Lyapunov equation, in the steady state, the diagonal and off-diagonal terms of the covariance matrix are related as follows;
\begin{eqnarray}
\mathcal{V}^s_{11}&=&\dfrac{\kappa}{2} \dfrac{1}{\kappa-\chi \cos(\phi)}+ \dfrac{\Delta_a + \chi \sin(\phi)}{\kappa-\chi \cos(\phi)}\mathcal{V}^s_{12} , \nonumber\\
\mathcal{V}^s_{22}&=& \dfrac{\kappa}{2} \dfrac{1}{\kappa + \chi \cos(\phi)}+ \dfrac{G}{\kappa+\chi \cos(\phi)} \mathcal{V}^s_{23} \nonumber\\
&-&\dfrac{\Delta_a - \chi \sin(\phi) }{\kappa+\chi \cos(\phi)}\mathcal{V}^s_{12},\nonumber\\
\mathcal{V}^s_{33}&=& \dfrac{2 n_b + 1}{2}+\dfrac{\omega_b}{\gamma}\mathcal{V}^s_{34}, \nonumber\\ 
\mathcal{V}^s_{44}&=&\dfrac{2n_b + 1}{2} + \dfrac{G}{\gamma}\mathcal{V}^s_{14}-\dfrac{\omega_b}{\gamma}\mathcal{V}^s_{34}.
\end{eqnarray}
Hence, the entropy production rate can be expressed using off-diagonal elements of the covariance matrix as 
\begin{widetext}
\begin{equation}
\Pi_s =  \dfrac{2\kappa \chi^2 \cos^2(\phi)}{\kappa^2_a-\chi^2 \cos^2(\phi)} + \dfrac{4\kappa \chi \cos(\phi) \left[\Delta_a+\kappa \tan(\phi)\right]}{\kappa^2-\chi^2 \cos^2(\phi)}  \mathcal{V}^s_{12} + \dfrac{2 G}{2n_b+1}\mathcal{V}^s_{14} + \dfrac{2 \kappa G}{\kappa+\chi \cos(\phi)}\mathcal{V}^s_{23}.
\label{entropy_prod2}
\end{equation}
\end{widetext}
Equation (\ref{entropy_prod2}) encompasses the full information on the role of optical parametric oscillator on the irreversibility of driven-dissipative optomechanical system. It shows that in the absence of the crystal nonlinearity $\chi\!=\!0$, the role of the correlations at the steady state is explicitly established~\cite{Brunelli2016ArXiv}. From Eq. (\ref{entropy_prod2}), even for small vanishing coupling $G\!=\!0$, the $\Pi_s$ is non-zero and depends explicitly on the contribution of the optical cavity mode. This is because the nonlinear interaction drives the system optical  mode into nonequilibrium state, which vanishes when $\chi\!=\!0$. However, it is also interesting to see that for finite nonlinear interaction, the $\Pi_s$ is modified by the contribution from both modes dynamical variables.
\begin{figure}[!t]
\includegraphics[width=0.98 \columnwidth]{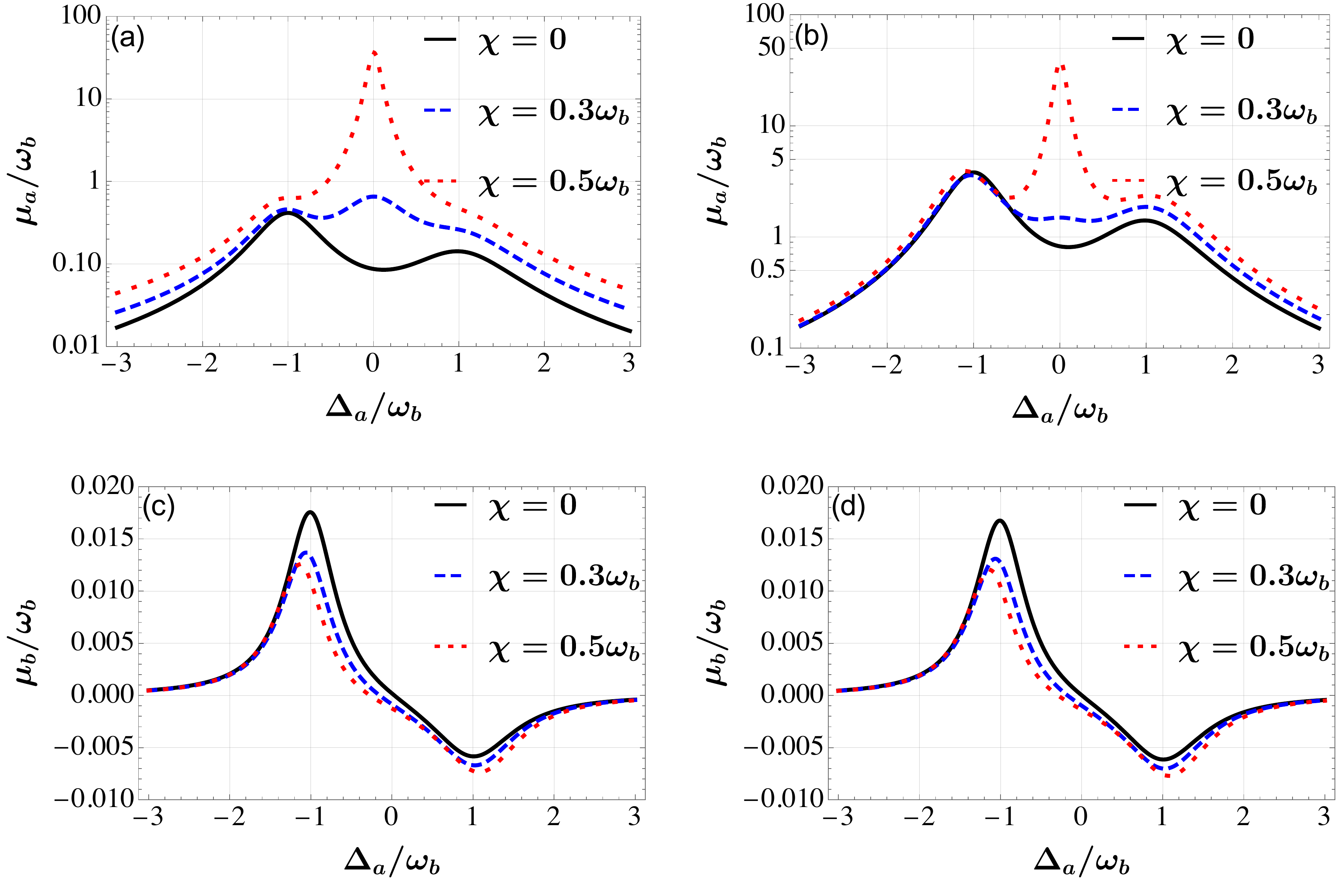} 
\caption{Scaled contributions $\mu_a/\omega_b$ and $\mu_b/\omega_b$ to the entropy production $\Pi_s$ against the normalized detuning $\Delta_a/\omega_b $ for different strength of the non-linear self-interaction of the mode $a$. The black solid curves corresponds to $\chi=0$, the dashed blue curves corresponds to $\chi=0.5\omega_b$ and the dotted red curves corresponds to $\chi=0.3\omega_b$. Panels (a) and (c) represent the plots when the number of thermal excitations $n_b=10$ while the panels (b) and (d) denotes the case of $n_b=100$. The other parameters are $\gamma=10^{-2}\omega_b$, $n_a=0$,  
$\kappa_a=0.5\omega_b$ and $G=0.1\omega_b$.
}          
\label{fig1}
\end{figure}

To numerically illustrate the influence of the non-linear contribution of the optical parametric oscillator placed in a cavity on entropy production at steady-state, we consider the resolved sideband regime $\kappa < \omega_b$.
In Fig. \ref{fig1} we present the  individual contributions $\mu_i$ ($i=a,b$) to the entropy production rate as a function of normalized detuning for different initial occupation of the cavity oscillator. In panels \ref{fig1}(a) and (b) [(c) and (d)], we plot the rescaled  cavity [atomic] contribution to the total entropy production rate  as a function of the detuning $\Delta_m/\omega_b$ for different values of $\chi$. For small value of coupling strength, $G=0.1\,\omega_b$, the system is stable in both the red-detuned region $\Delta_a >0$ and blue-detuned region $\Delta_a<0$. In addition, both contributions $\mu_a$ and $\mu_b$ are peaked at the two sidebands. In the limit of large detuning $\Delta_a \gg 1$, the two modes are effectively decoupled and leading to vanishing $\Pi_s$. It can be seen that the cavity  mode contribution to entropy production rate $\mu_a$ always increases as the non-linear self-interaction of the cavity mode $\chi$ increases. On the otherhand, Fig. \ref{fig1}(c) - (d), reveals sign changes in the mechanical mode component $\mu_b$ which captures the heating/cooling of the optomechanical system. We can see that $\mu_b$ shows a decreasing (increasing) behaviour for increasing $\chi$ in the blue-detuned (red-detuned) regime. For the contribution $\mu_a$, the behaviour appears symmetric between the red-detuned and the blue-detuned regimes with the  amplification being peaked at $\Delta_a=0$. 
From Fig. \ref{fig1} (a) and (b) we observe that the increase entropy production rate (irreversibility) associated with the presence of crystal nonlinearity $\chi\neq 0$ tends to vanish at $\Delta_a=1$ and $\Delta_a=-1$ as the number of thermal excitations increases. In comparable to the $\mu_b$, Fig. \ref{fig1} (c) and (d), there is no appreciable effect on the behavior of the entropy production rate outside the peaks for both change in $\chi\neq 0$ and the number of thermal excitations $n_b$.
\begin{figure*}[!ht]
\includegraphics[width=2 \columnwidth]{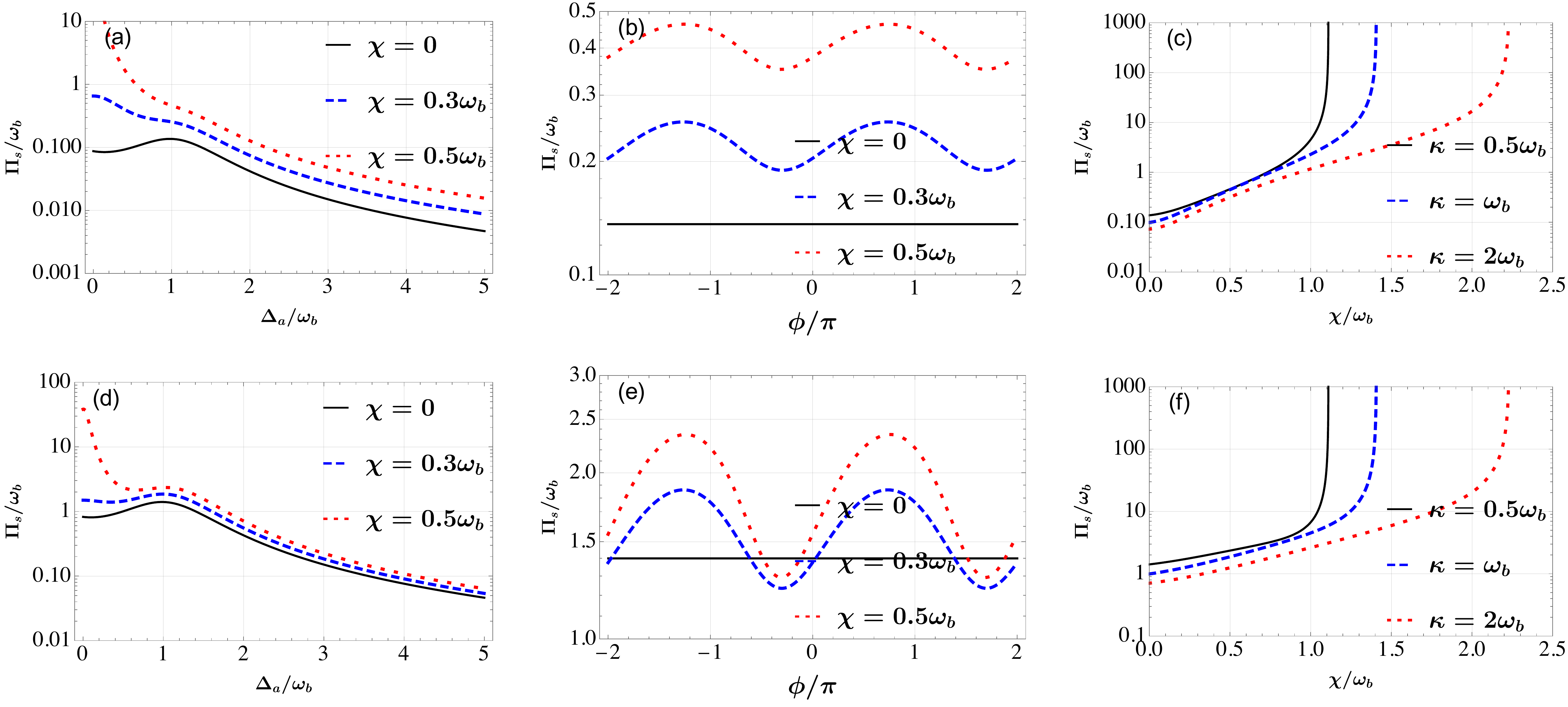} 
\caption{Entropy production rate $\Pi_s$ as function of normalized detuning $\Delta_a/\omega_b$ (first column) and  phase $\phi/\pi$ (second column) for different values of $\chi=0$ (solid black curve), $\chi=0.5 \omega_b $ (dashed blue curve) and $\chi= \omega_b$ (dotted red curve) where $\phi=0.7\pi$, $\gamma=10^{-2}\omega_b$, $n_a=0$,  $\kappa=0.51\omega_b$ and $G=0.1\omega_b$. (c) The third column, panels (c) and (f)), plot of $\Pi_s$ as function of strength of nonlinearity $\chi/\omega_b$ for different values of cavity decay rates $\kappa=0.5 \omega_b$ (black curve), $\kappa\!=\!\omega_b$ (blue curve) and $\kappa=2 \omega_b$ (red curve). The other parameters are $\gamma=10^{-2}\omega_b$,   $\phi=0.8\pi$ and $G=0.1\omega_b$. Panels (a), (b), (c) [the first row] represent the plots of thermal excitation  $n_b=10$ while panels (d), (e), (f) [second row] are the plots  for $n_b=100$. }  \label{fig2}
\end{figure*}

Fig. \ref{fig2} show the entropy production rate $\Pi_s$ against the rescaled detuning $\Delta_a/\omega_b$ and the phase $\phi$ for different values of non-linear parameter $\chi$ of the optical parametric oscillator as well as $\Pi_s$ against rescaled $\chi$ for different cavity decay rates $\kappa$. For the stable parameter range of the system, increasing the nonlinearity contribution enhances the entropy production rate.  In Fig. \ref{fig2} panel (a) - (c) we have considered a low  number of thermal excitation of the mechanical oscillator $n_b=10$  while the panel (d) - (f) assumed a very high initial occupation number $n_b=100$. Focusing on the red-detuned parameter regime, Fig. \ref{fig2} (a) shows that the entropy production rate $\Pi_s$ increases with the increasing  strength of crystal nonlinearity $\chi$ and diverges towards the resonance ($\Delta_a\!=\!0$). From plot (d), the case of high number of thermal excitation, it shows that the amount of irreversibility reduces but still finite, even when $\Delta_a\gg\omega_b$. Thus, the  initial thermal number of the excitation of the mechanical mode influences the $\Pi_s$ profile.

In Fig. \ref{fig2} (b) and (e), $\Pi_s$ are shown as a function of the phase $\phi/\pi$, for different values of $\chi$. Specifically plot (b) and plot (e) are calculated for lower and higher number of excitation respectively. We observe an increased irreversibility with a striking dips at $\phi\!=\! 1.7n\pi$ or $-0.25n\pi$ ($n$ is an integer)  due to phase of the non-linear self-interaction $\chi$ of the optomechanical system. In the sufficiently large initial number of thermal excitation $n_b=100$, the $\Pi_s$ for $\chi\!=\!0.5\omega_b$ dips below the absence of OPO case as shown in Fig. \ref{fig2} (e).  This clearly show that the irreversible entropy associated with the driving the system into the nonequilibrium state can be suppressed by choosing the appropriate phase-sensitive oscillator to incorporate into the an optomechanical system. Considering that the experimental optomechanical systems are affected by environmental noise, we present the impact of the cavity decay rate on the OPO modified setup in the third row of Fig. \ref{fig2} (i.e, (c) and (f)). It shows the robustness of the irreversibility associated with the OPO crystal non-linearity against the cavity decay rate $\kappa$. For small number of thermal excitation, $n_b=10$, the impact of the decay rate is minimal up to $\chi/\omega_b\simeq 1.0$ while the effect of increasing the cavity decay rate on $\Pi_S$ is more pronounce for very large number of thermal excitation. We remark that the observed reduction in the entropy production rate as the $\kappa$ increases and its linear grow with respect to $\chi/\omega_b$ is owing to the increase imbalance in populations between the two modes. We remark that the entropy production rate diverges at the point $\kappa^2\!=\!\chi^2\cos^2({\phi})$. 
\begin{figure*}[!ht] 
\includegraphics[width=2\columnwidth]{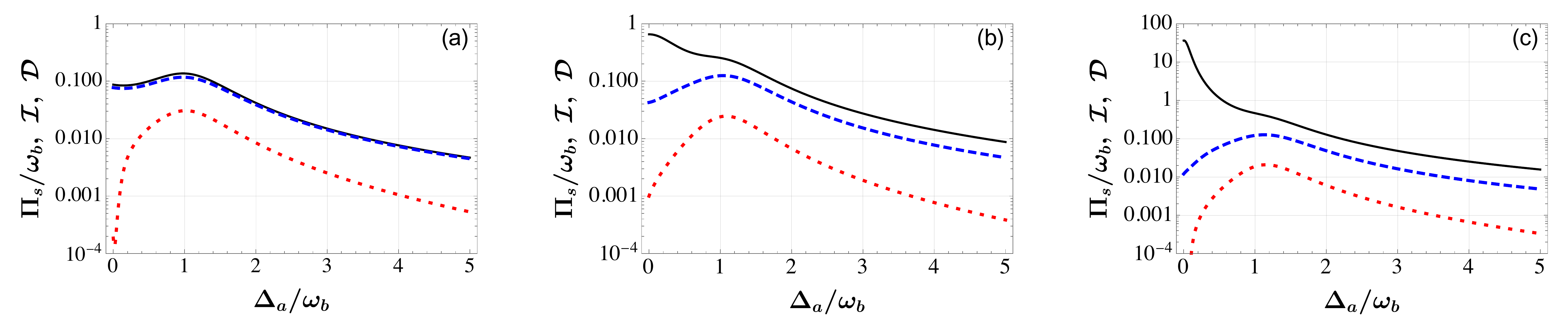} 
\caption{Entropy production rate $\Pi_s$ (solid black curve), mutual information $\mathcal{I}$ (dashed blue curve) and quantum discord $\mathcal{D}$ (dotted red curve) as function of normalized detuning $\Delta_a/\omega_b$ for different values of self non-linearity (a) $\chi=0$, (b) $\chi\!=\!0.3 \omega_b$ and (c) $\chi\!=\!0.5\omega_b$. The other parameters are $\phi=0.8\pi$, $\gamma\!=\!10^{-2}\omega_b$, $G=0.1\,\omega_b$, $\kappa\!=\!0.5 \omega_b$ and $n_b\!=\!10$.}
\label{fig3}
\end{figure*}
\subsection{Quantum correlations}\label{correlations}
Let now proceed to analyze how the presence of optical parametric oscillator in an optomechanical cavity influences the correlation profiles.
The net correlations between two modes can be quantified by means of the
quantum mutual information, 
\begin{equation}
    \mathcal{I}(\rho_{a:b}) = S(\rho_a) + S(\rho_b) - S(\rho_{ab}),
\end{equation}
where $S(\rho)=-\text{tr} \rho \ln \rho$ is the von Neumann entropy, and $\rho_a=\text{tr}_b\rho_{ab}$ and $\rho_b=\text{tr}_a\rho_{ab}$ are the reduced states of the two modes ($a$ and $b$). However, considering that the Gaussian nature of the states presented here, which are completely characterized by the two-mode convariance matrix $\mathcal{V}_{ab}$, it is more convenient to use the R\'enyi-2 entropy $S_2(\rho)=-\log \text{tr}[\rho^2]$. For a Gaussian state with covariance matrix $\mathcal{V}$, the R\'enyi-2 entropy can easily be evaluated and given by \cite{Adesso2012}
\begin{equation}
    S_2(\mathcal{V}) = \frac{1}{2}\ln (\text{det} \mathcal{V}).
\end{equation}
Thus, the Gaussian R\'enyi-2 mutual information for the two mode Gaussian state reads \cite{Adesso2012}
\begin{equation}
    \mathcal{I}(\mathcal{V}_{a:b}) = \frac{1}{2} \ln \left(\frac{\text{det} \mathcal{V}_a \, \text{det}\mathcal{V}_b}{\text{det} \mathcal{V}_{ab}}\right).
\end{equation}
Next we consider the measure of quantum discord based on the R\'enyi-2 entropy, which quantify the amount of quantum correlations beyond entanglement in Gaussian state. The quantum discord  is defined as the difference between the mutual information $\mathcal{I}(\mathcal{V}_{a:b})$ and the one way classical correlations $\mathcal{J}(\mathcal{V}_{a|b})$,
\begin{equation}
    \mathcal{D}(\mathcal{V}_{a|b}) = \mathcal{I}(\mathcal{V}_{a:b}) - \mathcal{J}(\mathcal{V}_{a|b}),
\end{equation}
where $\mathcal{J}(\mathcal{V}_{a|b})\!=\!\text{sup}_{\pi_b(X)}\{S(\mathcal{V}_a)-\int \text{d} X p_X S(\mathcal{V}^{\pi_b}_{a|X})\}$ is the maximum decrease in the R\'enyi-2 entropy of subsystem $a$, when a Gaussian measurement has been performed on subsystem $b$ such that $\pi_b(X)\!\ge\!0$, $\int \text{d}X \pi_b(X)\!=\! \mathbb{1}$ . 
Considering the maximization over all the possible measurements implemented on the mode $b$, we can express as
\begin{equation}
     \mathcal{D}(\mathcal{V}_{a|b})\!=\!\frac{1}{2}\ln (\text{det}\mathcal{V}_b) - \frac{1}{2}\ln (\text{det} \mathcal{V}_{ab}) + \text{inf}_{\pi_b} \frac{1}{2}\ln (\text{det}\mathcal{V}_a^{\pi_b}).
\end{equation}

It have recently  been demonstrated that the irreversibility generated by the steady state and the total amount of correlations shared between two coupled oscillators are closely related \cite{Brunelli2016ArXiv}.
In what follows, we focus on the mutual information and quantum discord between the two modes at the stationary state as well as the influence of self non-linear interaction on them.
In Figure \ref{fig3} we compare the entropy production rate $\pi_S$ to the correlations established by the optomechanical system with a driven nonlinear crystal, as quantified by the mutual information $\mathcal{I}$ and quantum discord $\mathcal{D}$ at the phase $\phi\!=\!0.8\pi$. In panel (\text{a}), for $\chi\!=\!0$, we see a close similarity between the entropy production rate and the mutual information curves. For $\chi\ne 0$, Fig. \ref{fig3} (b) and (c), there is a striking difference between the entropy production rate and the quantum correlations. It can be seen that the entropy production $\Pi_s$ increases when $\Delta_a/\omega_b<0$ while the mutual information $\mathcal{I}$ and discord $\mathcal{D}$ are decreasing. The deviation can be attributed to the modification of the cavity decay rate by the nonlinear medium.

\section{Conclusions}\label{conclusions}
We have studied the irreversible entropy generated in an interacting nonlinear hybrid optomechanical cavity system by stationary driven dissipation process.   We have investigated the scenario in which a non-linear medium is placed inside a driven optomechanical system. The system can be well described by two mode composite Gaussian system. We have shown that the stationary state entropy production rate depends on the strength of the nonlinear self interaction of the optical cavity mode. Our analysis showed that the presence of the nonlinear crystal decreases the entropy production rate and quantum correlations  as nonlinearity increases. We have further shown that the relationship between the entropy production rate and the  quantum correlations are drastically modified by the nonlinear medium.
We remark that our investigation can easily be implemented in the current state-of-art experimental technology. As reported in optomechanical setup experiment \cite{Pirkkalainen2015},  the frequency of cavity $\omega_C=2\pi \times 4.93$ GHz, the decay rate of the cavity  $\kappa\!=\!2\pi \times 215$ KHz, the frequency of mechanical resonator  $\omega_b\!=\!2\pi\times 65$ MHz,  the corresponding mechanical damping rate  $\gamma\!=\!2\pi 15$ KHz, and the single-photon optomechanical coupling strength  $g\!=\!2\pi\times 1.6$ MHz.  Our work would benefit the current effort towards optimization of quantum thermal devices \cite{Myers2022,Sheng2021,Naseem2019} and the better understanding energetic cost of cooling optomechanical systems \cite{Monsel2021}.
\section*{Acknowledgements}
MA and BT were supported by Khalifa University through project no.8474000358 (FSU-2021-018).
COE and NA were supported by LRGS Grant LRGS/1/2020/UM/01/5/2 (9012-00009)  provided by the Ministry of Higher Education of Malaysia (MOHE). OA acknowledged the Newcastle University Academic Track Fellowship.

%

\end{document}